\documentclass{sf2a-conf2010}
\usepackage{amsmath}
\usepackage{graphicx}
\usepackage{hyperref}
\usepackage[]{natbib}

\newcommand{\eps}{\varepsilon}
\newcommand{\Msun}{\ensuremath{M_\odot}}
\begin{document}
\TitreGlobal{SF2A 2010}
%
\title{Galaxy Formation: Merger vs Gas Accretion}
\author{B. L'Huillier}\address{LERMA, Observatoire  de  Paris, UPMC,
  61 avenue de l'Observatoire, 75014 Paris, France}
\author{F. Combes$^{1}$}
\author{B. Semelin$^{1}$}
\runningtitle{Merger vs Gas Accretion}
%
\setcounter{page}{237}
\index{L'Huillier, B.}
\index{Combes, F.}
\index{Semelin, B.}

\maketitle
\begin{abstract}
%
%
According  to the hierarchical  model, small  galaxies form  first and
merge together to form bigger objects. 
In  parallel,  galaxies assemble  their  mass  through accretion  from
cosmic filaments. 
Recently,  the  increased   spatial  resolution  of  the  cosmological
simulations have emphasised  that a large fraction of  cold gas can be
accreted by galaxies. 
In  order  to   compare  the  role  of  both  phenomena  and  the
corresponding star formation history, one has to detect the structures
in the numerical simulations and to follow them in time, by building a
merger tree.
\end{abstract}
\section{Introduction}

Recent   simulations (\citet{kere_do_2005}  for   example)  have
emphasised the role  of smooth cold accretion on  galaxy formation. We
aim  at comparing the  roles of  mergers and  gas accretion  on galaxy
growth by studying numerical simulations.

\section{The Simulations}
We       use       a        set       of       TreeSPH       multizoom
simulations~\citep{semelin_formation_2002,semelin_new_2005},   starting
with  a  low  resolution  cosmological  simulation,  and  resimulating
regions of interest at higher resolution. 
%
The box radius is 8.30~Mpc (comoving), the mass resolution is $3\times
10^7$~\Msun{}  for   baryons  and  $1.4\times   10^8$~\Msun{}  for  DM
particles, and the gravitational smoothing is
$\eps = 6.25$~kpc.  
There  are 90  outputs spaced  by 100~Myr  from $z\sim{29}$  to $z\sim
{0.41}$, which enables us to follow particles from one output to the other.


\section{Structure Detection and Merger Tree}

We  use  AdaptaHOP \citep{aubert_origin_2004,tweed_building_2009}  to
detect the DM haloes and subhaloes hierarchy.  
We also use AdaptaHOP to detect the baryonic galaxies \footnote{Screenshots of the
  simulations and of  the structures found by AdaptaHOP  as well as an
  example      of     merger     tree      can     be      seen     at
  \url{http://aramis.obspm.fr/~blhuill/research.html}},  with a better
adapted set of parameters: 
we use for  the density threshold above which  structures are detected
$\rho_\text{T} =  {1000}$ (times the  mean density of  the simulation)
instead of 81 for  DM. We also check that the results  do not vary too
much with the choice of $\rho_\text{T}$.

In the following, we are only interested in baryonic particles and structures.
At  each timestep,  baryonic particles  either  belong to  a structure
(galaxy or satellite), or are diffuse and belong to the background.
To compute  the mass gained  by the main  galaxy at each  timestep, we
sum the  mass of all the  particles entering the  structure, and we
count  as  \emph{smooth  accretion}  particles that  belonged  to  the
background at  the previous  timestep, and as  \emph{merger} particles
that belong to another structure. 
Particles  can  also leave  the  main  galaxy  for another  structure,
generally for a satellite (\emph{fragmentation}) or for the background
(\emph{evaporation}). 
We  then   have  to  substract  \emph{fragmentation}   to  \emph{merger}  and
\emph{evaporation} to \emph{accretion}.

One of the main problems while building a merger tree is the so-called
\emph{flyby issue}: when structures are too close one to another, they
are undistinguishable for the structure finder.
Thus two structures can be separated at a given timestep, merged at the
following, and separated again later.
Such an example can be seen in figure~\ref{fig:accretion}, left: the red upper curve
(satellites) grows, then decreases  when the satellite flies away from
the central galaxy, then the curve increases again when the satellite comes back.

Thus with  our technique, when  particles enter the main  galaxy, they
are counted positively, and  negatively when they leave, which enables
us   to  compute   the  total   mass   origin  of   the  main   galaxy
(figure~\ref{fig:accretion}, right).

\section{Results}
\begin{figure}[h]
  \begin{center}$
    \begin{array}{cc}
      \includegraphics[width=8cm]{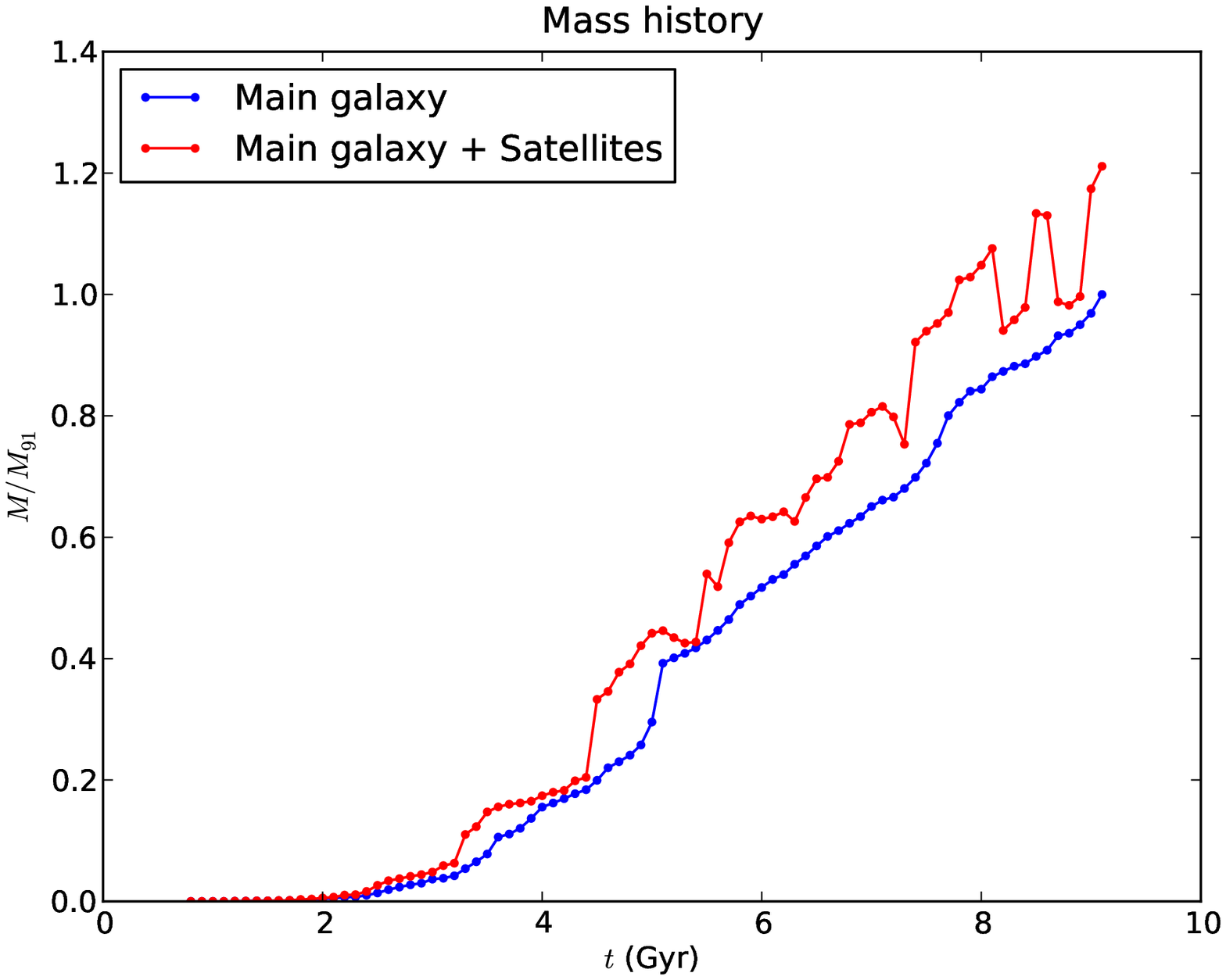} & 
      \includegraphics[width=8cm]{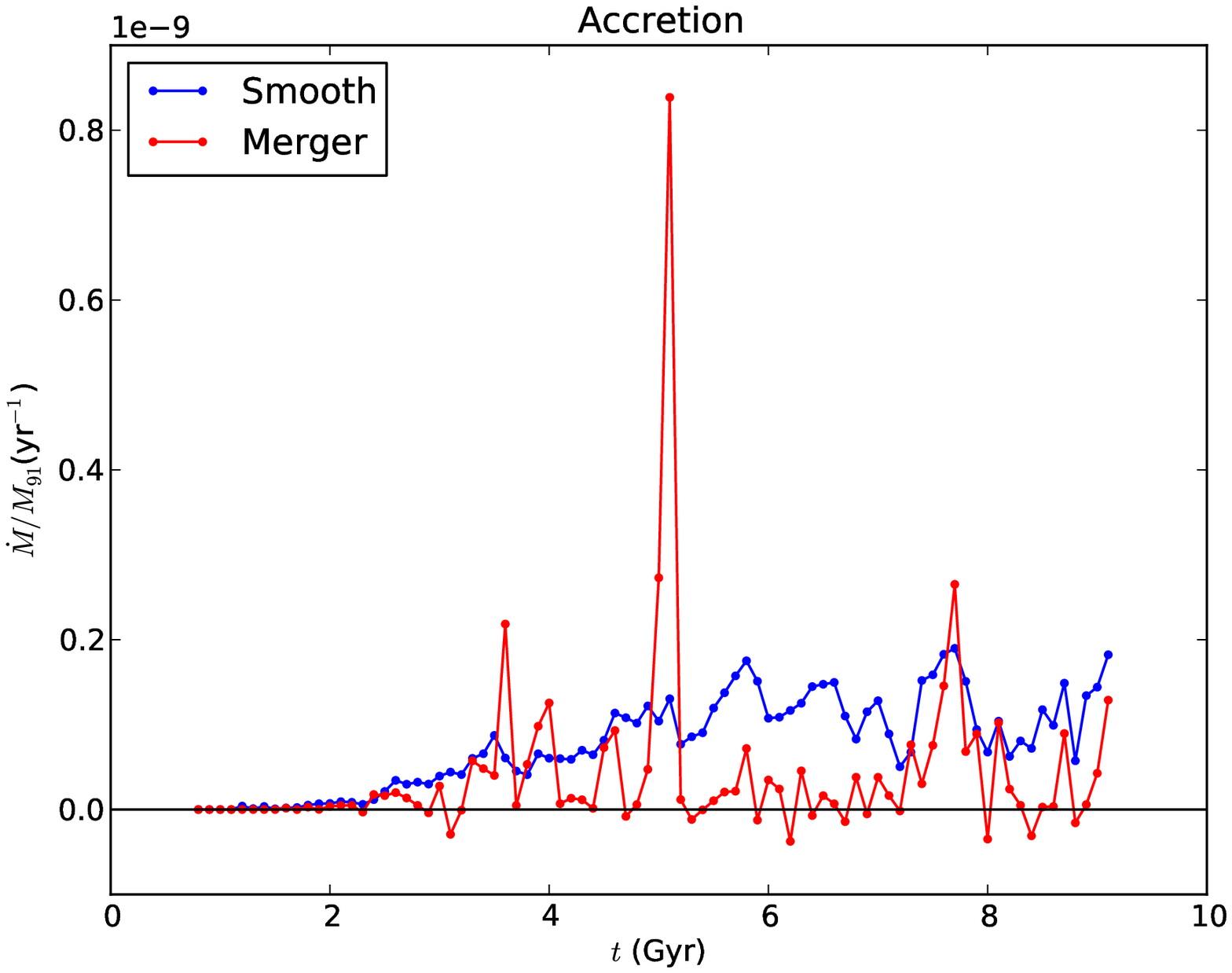}\\
    \end{array}$
    \caption{\label{fig:accretion}    \emph{Left:}    Baryonic    mass
      evolution of a galaxy 
      normalised by the galaxy mass at $t = 9.1$ Gyr. 
      \emph{Right:} Incoming mass by time unit.  The
      blue curve  shows mass smoothly  accreted, the red  curve shows
      the mass gained through mergers.}
  \end{center}
\end{figure}

We measure  the smooth accretion  and the merger fractions  of several
galaxies, as shown in figure~\ref{fig:accretion}, right and table~\ref{tab:res}.
\vspace{-1cm}
\begin{table}[ht]
  \caption{\label{tab:res}Fraction of  smooth accretion within  the total assembled
    mass  for  different  central   galaxies.  The  first  galaxy  has
    $f_\text{acc} >  1$, which means  that the galaxy loses  more mass
    during merger envents due to fragmentation that it gains.}
  \centering
  \begin{tabular}{|l|l|l|l|l|l|l|}
    \hline
     galaxy &1 & 2&3&4&5&6\\
    \hline
    Mass ($10^{11} \Msun$) & 107.5 &244.81 & 140.81 & 1.73 &143.40& 8.98 \\
    \hline
    Accretion fraction      & 1.04$^*$   & 0.65  & 0.67   & 0.52 & 0.95 & 0.71   \\
    \hline
   
  \end{tabular}
\end{table}
		

\section{Conclusions}
The study  of these simulations  shows that baryonic mass  assembly of
galaxies seems to be dominated  by smooth accretion, although we still
have to perform further consistency tests.
The  next  step is  to  perform  statistical  studies to  confirm  the
preliminary results,  then further  physical exploitation can  be made
such as the role of the environment on the SFR.

\begin{acknowledgements}
BL would like to thank D. Tweed and S. Colombi for stimulating discussions.
\end{acknowledgements}

\bibliographystyle{aa}
\bibliography{thesis}

\end{document}